\documentclass[%
reprint, aps, pre,
 amsmath,amssymb,
]{revtex4-2}

\usepackage{graphicx}
\usepackage{dcolumn}
\usepackage{bm}

\newcommand{\vc}{\boldsymbol}

\begin{document}

\preprint{}

\title{Note on the physical basis of spatially resolved thermodynamic functions}

\author{Rasmus A. X. Persson}
 \email{rasmus.persson@gu.se}
\affiliation{Department of Pedagogical, Curricular \& Professional Studies, University of Gothenburg, Gothenburg, Sweden}%

\date{\today}

\begin{abstract}
The spatial resolution of extensive thermodynamic functions is discussed. A physical definition of the spatial resolution based on a spatial analogy of partial molar quantities is advocated, which is shown to be consistent with how hydration energies are typically spatially resolved in the molecular simulation literature. It is then shown that, provided the solvent is not at a phase transition, the spatially resolved entropy function calculated by first-order grid inhomogeneous solvation theory (Nguyen et al. J. Chem. Phys., 137, 044101 [2012]) satisfies the definition rigorously, whereas that calculated by grid cell theory (Gerogiokas et al., J. Chem. Theory Comput., 10, 35 [2014]) most likely does not. Moreover, for an ideal gas in an external field, the former theory is shown consistent in the limit of weak field or high temperature whereas the latter is not. Finally, consistent with the proposed definition and with the case of an ideal gas in an external field, we derive an approximate expression for the solvent contribution to the free energy of solvation in the limit of infinite dilution from the spatial variation of the density around the solute.
\end{abstract}

\maketitle


\section{Introduction}
\label{sec:intro}
We consider an arbitrary extensive thermodynamic quantity $A$, say the energy or the entropy. In experiments, typically only changes in $A$ as the system undergoes transformations can be inferred. For model systems, statistical mechanics often allows a calculation of $A$, either analytically in some special cases or numerically with the aid of computers using the methods of molecular simulation or integral equation theories.

The calculation of $A$ through statistical mechanics naturally involves variables that are hidden in the macroscopic experiment. Mark and van Gunsteren \cite{mark94} criticized any attempt at decompositions of the entropy (or free energy) in terms of particular atomic interactions for non-ideal systems. Nevertheless, many
authors \cite{nguyen12,raman13,gerogiokas2014prediction,nguyen14,michel14,gerogiokas15,nguyen15,raman15,persson2017signatures,heinz2019computing,heinz2021per,waibl2022explicit} have studied decompositions of the form (in this Note, all integrals are to be interpreted as definite integrals over the entire liquid phase),
\begin{equation}
\label{eq:decomposition}
A = \int \mathrm d^3 \vc x \overline A(\vc x)
\end{equation}
where $\overline A(\vc x)$ is the \emph{local} density of $A$ at
position $\vc x$ in the system. Since $A$ is an extensive quantity, such a decomposition appears natural. The problem is that the conclusions drawn from an analysis of the spatial behavior of $\overline A(\vc x)$ depend on the choice of this function, as it is not uniquely defined by eq.~(\ref{eq:decomposition}) alone.

To stress the main point of this Note: in theory \emph{any} spatial decomposition of an extensive thermodynamic function that is consistent with the integrated (macroscopic) value can be obtained and \emph{none} can be said to be more ``correct'' than the other from an experimental point of view. In other words, any conclusion can in principle be drawn from an analysis of spatially resolved thermodynamic functions without there being, even in theory, any rigorous way to discern ``good'' from ``erroneous'' conclusions.

It is certainly the case that the approximation that the use of atomic or molecular force fields entails means that also non-statistical properties such as the energy become approximate when expressed in truncated fashion as a sum of two-body terms. This is an error, or inconsistency, with respect to ``experiment'' where the energy is actually resolved over atomic nuclei and electrons, but the caution that Mark and van Gunsteren \cite{mark94} raise regards an inconsistency  \emph{within} the model itself. In other words, even if the force field were exact (taking into account the internal atomic degrees of freedom), decomposition of the free energy or of the entropy over specific atomic interactions or spatial points would still not be uniquely defined.

As a more minor point, we shall appeal to the theory of mixtures and seek to introduce a general axiomatic form for $\overline A(\vc x)$. For simplicity, our starting point will be a solvation process of a monoatomic solute, so that changes in $A$ correspond to changes in the solvent only. Our definition for $\overline A(\vc x)$ will be shown to coincide with how the energy is usually decomposed spatially in the literature. Consistency dictates that the entropy should be spatially resolved in the same manner, not the least to ensure the correct interpretation of any entropy-energy compensation \cite{ben1975hydrophobic}.

\section{Spatially resolved thermodynamics of solvation}
\label{sec:solv}
Provided with eq.~(\ref{eq:decomposition}), it follows that the total change in $A$ in an arbitrary solvation process is given by,
\begin{equation}
\label{eq:solvation}
\Delta A = \int \mathrm d^3 \vc x [\overline A(\vc x) - \overline
A_\infty ]
\end{equation}
where $\overline A_\infty$ is the $A$-density function evaluated for the bulk solvent (infinitely far removed from the solute). Whereas eq.~(\ref{eq:solvation}) follows from eq.~(\ref{eq:decomposition}) in total, we impose the extra condition (``meanfield approximation'') that the integrand $\overline A(\vc x) - \overline A_\infty$ equals the \emph{local} change in $A$ when the solvent is restructured at point $\vc x$ from its initial (unperturbed) bulk state to its (perturbed) solvation state. If, for simplicity, we restrict ourselves to a monoatomic solvent, this local change can only manifest itself through the local fluid density. We therefore seek to express $\overline A(\vc x)$ as a function of the local density, or particle number, only and ignore any dependencies on derivatives or spatial correlations of this quantity. As we shall see in Section~\ref{sec:energy}, the definition which we shall present coincides with what is commonly implicitly used for the energy (enthalpy) \cite{nguyen12,gerogiokas2014prediction,gerogiokas15,persson2017signatures,waibl2022explicit}.

Because of correlations, the individual volume elements of the fluid are not
independent subsystems. If they had been, the extensivity of $A$ would
imply a linear dependence on the number of molecules in the subsystem---at
\emph{fixed} pressure $P$ and temperature $T$---and $\overline A(\vc x) v(\vc x)$ could be identified with the change in $A$ upon addition of one molecule in a small volume $v(\vc x)$ centered at $\vc x$. For interacting subsystems, this relation is not necessarily linear, and we therefore make an infinitesimal change, so that the local
$A$ at position $\vc x$ is $$\left (\frac {\partial A}
{\partial N(\vc x)} \right )_{T,P},$$ and hence $\overline A(\vc x)$ of eq.~(\ref{eq:decomposition}) becomes:
\begin{equation}
\label{eq:def}
\overline A(\vc x) = \rho(\vc x) \left ( \frac {\partial A}
{\partial N(\vc x)} \right )_{T,P}
\end{equation}
Here $\rho(\vc x)$ and $N(\vc x)$ denote, respectively, the number density and
number of molecules at position $\vc x$, both of
which are fluctuating quantities: appropriate ensemble averaging is implicitly
understood for both. By the derivative with respect to the number of particles, we actually understand the finite change when adding one particle to the system. The use of the derivative symbol in this case is normally justified by the number of particles in a thermodynamic system being very large. While this argument is not applicable here since the ensemble averaging entails that $N(\vc x)$ is -- in the technical jargon of statistics -- ``almost surely'' zero, it does not deter us and we simply regard the derivative symbol as a notational shorthand for a finite difference. 

That $\overline A(\vc x)$ satisfies eq.~(\ref{eq:decomposition}) (in the sense that its spatial integral yields $A$) can be seen the most easily by using a discrete approximation for $\rho(\vc x)$ in which space
is divided into $M$ cells, indexed by $i$, of a small but finite volume $v$. Writing
\begin{equation}
\label{eq:def-discrete}
\overline A_i = \frac {N_i} {v} \left (\frac {\partial A}
{\partial N_i} \right )_{T,P}
\end{equation}
where $N_i$ is the number of molecules in cell $i$, we have the
differentials
\begin{equation}
\mathrm d A = \sum_i^M \left (\frac {\partial A}
{\partial N_i} \right )_{T,P} \mathrm d N_i
\end{equation}
and
\begin{equation}
\mathrm d \overline A_i = \frac {\mathrm d N_i} {v} \left ( \frac {\partial
A} {\partial N_i} \right )_{T,P}
\end{equation}
which combined lead to a discrete analog of eq.~(\ref{eq:decomposition}), \textit{viz.}
\begin{equation}
    A = \sum_i^M v \overline A_i.
\end{equation}

Two remarks are justified before we continue. First, the quantity $(\partial A / \partial N(\vc x))_{T,P}$ is the partial molar $A$ of solvation for a solvent molecule kept fixed at position $\vc x$ (this position is defined with respect to the location of the solute); this is a theoretical device by which diffusion of the solvent in the liquid is taken to be analogous to chemical interconversion, each solvent molecule being defined not only by its chemical identity but also by its physical location. For the bulk fluid, this expression then equals the partial molar $A$ of solvation in itself \textit{tout court}  because of translational symmetry in the fluid. Only in the inhomogeneous fluid (around a solute) will there be a spatial dependence (solvent molecules close to a solute will behave differently from those far in the bulk). Second, in the particular case that $A$ is the Gibbs energy, then the partial derivative in question is the chemical potential which is also constant throughout the fluid, whether homogeneous or not.

\section{Spatial decompositions in the literature}
Before continuing the development in Section~\ref{sec:consistent}, we here digress slightly to see whether some of the spatial decompositions proposed in the literature conform with the axiomatic approach above. We deal with the spatial resolution of the energy and of the entropy separately for reasons that will be clear shortly.

\subsection{Energy}
\label{sec:energy}
In molecular simulations, the total potential energy may be seen as the sum of solvent-solvent $U_\mathrm{ss}$,  solute-solvent $U_\mathrm{xs}$ and kinetic contributions. These last ones are entirely local and thus basically ``resolve themselves'' spatially. The first two contributions (the potential energy) are computed as a sum over individual terms over all of the molecules. This leads to a very intuitive scheme for their spatial decomposition, in that the energy terms involving molecules are projected onto the corresponding positions in which said molecules are located. For the solvent-solvent contributions, this means
\begin{equation}
\overline U_\mathrm{ss}(\vc x) = \frac 1 2 \left \langle \sum_{i} \delta(\vc r_i - \vc x) \sum_{j \neq i}  u_\mathrm{ss}(|\vc r_j - \vc r_i|) \right \rangle 
\end{equation}
where $\delta(\vc r)$ is the three-dimensional Dirac delta-function, $u_\mathrm{ss}(r)$ is the pair potential between solvent molecules and the sums run over all molecules positioned at $\{\vc r_i\}$ and the angle brackets denote proper ensemble averaging over all positions $\{\vc r_i\}$. In the averaging procedure, the delta-function picks out only those contributions for which $\vc x$ coincides with a molecular center. In other words, the energy associated with spatial position $\vc x$ is taken as the average energy required to remove a molecule from position $\vc x$ to infinity, and the factor of $1/2$ corrects for double counting \footnote{Including this factor leads to the spatial decomposition of the \emph{average} energy per molecule; otherwise it leads to that of the total \emph{binding} energy: the energy difference of the system with respect to its isolated constituent molecules. In the literature, it is more common to see the former spatially resolved than the latter, but it is a subtle question which one is actually more appropriate for local thermodynamics.}. Since only terms with $\vc r_i = \vc x$ contribute to the summations, the sum over $j$ becomes independent of the sum over $i$, and the expression can be rewritten as \footnote{Whether one excludes from the average, terms for which $\vc r_j = \vc x$, or simply takes $u_\mathrm{ss}(0)$ finite by definition, does not affect the converged ensemble average.}
\begin{equation}
\overline U_\mathrm{ss}(\vc x) = \frac 1 2 \left \langle \sum_{i} \delta(\vc r_i - \vc x) \right \rangle \left \langle \sum_{j} u_\mathrm{ss}(|\vc r_j - \vc x|) \right \rangle.
\end{equation}
Hence, recalling that $\rho(\vc x)$ is the ensemble-averaged density, we have
\begin{equation}
\overline U_\mathrm{ss}(\vc x) = \frac 1 2 \rho(\vc x) \left \langle \sum_{i} u_\mathrm{ss}(|\vc r_i - \vc x|) \right \rangle
\end{equation}
and eq.~(\ref{eq:consistent-energy}) is thus seen to satisfy eq.~(\ref{eq:def}) up to an error, inversely proportional to the total system size, that disappears in the thermodynamic limit \footnote{Since the relative fluctuations of the energy are inversely proportional to the \emph{square-root} of the total system size, the error in eq.~(\ref{eq:consistent-energy}) is negligible next to the thermal fluctuations already for systems of a few thousand molecules such as in molecular simulations.}.

The solute-solvent contribution is decomposed in the same way,
\begin{equation}
\overline U_\mathrm{xs}(\vc x) = \left \langle \sum_{i} \delta(\vc r_i - \vc x) u_\mathrm{xs}(|\vc x|) \right \rangle 
\end{equation}
where the monoatomic solute is taken to lie at the origin of the coordinate system and $u_\mathrm{xs}$ is the solute-solvent pairwise interaction energy. This leads to
\begin{equation}
\overline U_\mathrm{xs}(\vc x) = \rho(\vc x) \left \langle u_\mathrm{xs}(|\vc x|) \right \rangle.
\end{equation}
 The spatial decomposition of the total potential energy is simply the sum of these two contributions,
\begin{equation}
    \label{eq:consistent-energy}
    \overline U(\vc x) = \overline U_\mathrm{ss}(\vc x) + \overline U_\mathrm{xs}(\vc x)
\end{equation}
and this is essentially the procedure used in Refs.~\onlinecite{nguyen12,gerogiokas2014prediction,gerogiokas15,persson2017signatures,waibl2022explicit} and many others, and it seems to be the only intuitive one. The kinetic energy density, in turn, is simply given by $\frac 3 2 T \rho(x)$ for a monoatomic solvent, when taking Boltzmann's constant to be unity, a convention that we shall observe throughout.

\subsection{Entropy}
There is no intuitive spatial resolution of the entropy as there is for the intermolecular energy. Consequently, there are different suggestions in the literature, of which we shall briefly comment on three. In Section~\ref{sec:consistent}, we shall suggest a fourth alternative.

\subsubsection{Grid cell theory}
We now consider whether the entropy of grid cell theory
\cite{gerogiokas2014prediction} (GCT; a spatially resolved variant
of Henchman's cell theory \cite{henchman03}) conforms to eq.~(\ref{eq:def}), in order to be consistent with how the energy is resolved in the cited reference. In this theory, the configuration space is
divided into discrete cells and the average magnitude of the force of every
molecule within a specific cell is computed. The local entropy density of the
cell is then computed as \footnote{For notational simplicity, we
only consider the translational contribution, as we have only dealt with monoatomic molecules elsewhere.},
\begin{equation}
\overline S_i^\mathrm{GCT} = \frac {N_i} {v} S^\mathrm{HO}_i
\end{equation}
where $S^\mathrm{HO}_i$ is the entropy of the harmonic oscillator whose average
force magnitude $f_i$ equals that of the molecules in cell $i$:
$S_i^\mathrm{HO} \propto \ln f_i$. This choice of matching the molecule of the fluid to the harmonic oscillator is arbitrary and there is no \textit{a priori} reason to prefer matching the average force magnitude to, say, the average square of the force. However, the choice will impact the calculated spatial entropy distribution. This entails that if the GCT entropy should satisfy definition (\ref{eq:def}) for one particular matching choice, it will not for another one.

Since the total entropy is taken as [to satisfy eq.~(\ref{eq:decomposition})]
\begin{equation}
S_\mathrm{tot}^\mathrm{GCT} = \sum_j^M N_j S_j^\mathrm{HO}
\end{equation}
then $S^\mathrm{HO}_i = ({\partial S_\mathrm{tot}^\mathrm{GCT}} / {\partial
N_i})_{T,P}$ only if
\begin{equation}
\label{eq:cond}
\sum_{j \neq i} N_j \left (\frac {\partial S_j^\mathrm{HO}} {\partial N_i}
\right )_{T,P} = -N_i \left (\frac {\partial S^\mathrm{HO}_i} {\partial N_i}
\right )_{T,P}
\end{equation}
which using the chain rule for derivatives, simplifies to
\begin{equation}
\label{eq:cond-force}
\sum_{j \neq i} \frac {N_j} {f_j} \left (\frac {\partial f_j} {\partial N_i}
\right )_{T,P} = -\frac {N_i} {f_i} \left (\frac {\partial f_i} {\partial N_i}
\right )_{T,P}
\end{equation}
where $f_k > 0$ is the average \emph{magnitude} of the molecular forces in cell $k$
and $N_k > 0$ is the corresponding number of
molecules. This condition expresses a peculiar, and possibly unphysical, condition in that, up to a numerical factor which is unity for the homogeneous fluid, the average change in magnitude of the molecular forces in all other cells should always exactly cancel the corresponding change in the cell where a molecule is added. Of course, for the homogeneous fluid, all derivatives in eq.~(\ref{eq:cond-force}) vanish and the equation is satisfied trivially. Seeing as any interesting spatial resolution is predicated on the fluid being inhomogeneous around a solute, we shall not pay further attention to this case.

Let us out of simplicity examine this equation in the special case of a system
that is divided into two grid cells that together contain all of the molecules
in the system. Eq.~(\ref{eq:cond-force}) may then be cast as
\begin{equation}
\label{eq:force}
\frac {N_2} {f_2} \left (\frac {\partial f_2} {\partial N_1} \right )_{T,P} =
- \frac {N_1} {f_1} \left ( \frac {\partial f_1} {\partial N_1} \right )_{T,P}
\end{equation}
Clearly, this amounts to a requirement 
that the change in the average \emph{magnitude} of the molecular force in one cell is opposite in sign to the corresponding change in the other cell when a molecule is added to either. Since the average total force is zero, eq.~(\ref{eq:force}) implies that the distribution of molecular forces should narrow in one cell if it broadens in the other one, when adding a molecule. While this condition is very stringent, it is difficult to prove explicitly that it is not fulfilled by GCT short of performing numerical simulations.

The inconsistency of GCT is more readily proved in a different way. Consider an ideal gas in an external field, so that the gas density is inhomogeneous, an idealized model of the Earth's atmosphere. Now
\begin{equation}
    \left ( \frac {\partial f_k} {\partial N_i} \right )_{T,P} = 0
\end{equation}
for arbitrary $i$ and $k$ since for the ideal gas the force depends only on the external field and not on the other molecules. If we for simplicity assume that the field is homogeneous (as it will be over sufficiently short distances), then the local entropy of GCT becomes a spatial constant. However, this means that the ``gravito-chemical'' potential (in the continuum limit),
\begin{equation}
    \mu(\vc x) = \langle u_\mathrm{xs}(|\vc x|) \rangle + \frac 3 2 T - T S^\mathrm{HO}
\end{equation}
is not constant, since the first term depends on $\vc x$ but the other ones do not. This is in clear contradiction with the postulates of equilibrium thermodynamics.

\subsubsection{Grid inhomogeneous solvation theory}
While not based directly on eq.~(\ref{eq:solvation}), it is nevertheless
instructive to investigate to what extent the spatially resolved solvation
entropy, computed by first-order grid inhomogeneous solvation theory (GIST) in
Ref.~\onlinecite{nguyen12}, satisfies eq.~(\ref{eq:def}). Note that GIST provides equations for the change of entropy in the solvation process, whereas GCT provides equations for the absolute entropy. This difference is immaterial to our arguments, since the pure solvent entropy is a spatial constant in any case.

In the present notation, the first-order local GIST solvation entropy density is written \footnote{Only the translational contribution is considered.},
\begin{equation}
\label{eq:something}
\Delta \overline S_i^\mathrm{GIST} = -\frac {N_i} {v} \ln
\left ( \frac {N_i} {v \rho_\infty} \right )
\end{equation}
where $\rho_\infty$ is the bulk density, and so ($v$ implicitly depends on
$N_i$ due to the condition of constant pressure)
\begin{equation}
\left (\frac {\partial \Delta S_\mathrm{tot}^\mathrm{GIST}} {\partial N_i}
\right )_{T,P}  = -\left [ \ln \left (\frac {N_i} {v \rho_\infty} \right ) + 1
- \frac {N_i} {v} \left (\frac {\partial v} {\partial N_i} \right )_{T,P}
\right ]
\end{equation}
Multiplying this result by $N_i / v$ we do not recover eq.~(\ref{eq:something}) as we should by eq.~(\ref{eq:def-discrete}), and so this proves that eq.~(\ref{eq:def}) is not satisfied in general save for an ideal gas
(for which $(\partial v / \partial N_i)_{T,P} = v / N_i$ always). However, in the single-phase region (outside the binodal line, \textit{i.~e.} no phase transitions), eq.~(\ref{eq:def}) \emph{is} satisfied also for a non-ideal solvent, because the molecular volume is then uniquely defined by the thermodynamic state which does not change upon molecule insertion (no singularity of the compressibility in the thermodynamic limit). This is far from presenting any actual limitation when it comes to biomolecular simulations of aqueous solutions and GIST is thus arguably superior in this respect compared to GCT.

For completeness, we consider also the case of the ideal gas in the external field. In the case of first-order GIST, we have,
\begin{equation}
\label{eq:gist}
    \mu(\vc x) = \langle u_\mathrm{xs}(|\vc x|) \rangle + \frac 3 2 T + T \ln \left ( \frac {\rho(\vc x)} {\rho_\infty} \right )
\end{equation}
in the continuum limit. But clearly,
\begin{equation}
\ln \rho(\vc x) = \rho_\infty \left \langle \exp \left (-\frac {u_\mathrm{xs}(|\vc x|)} T \right ) \right \rangle.
\end{equation}
This means that first-order GIST is consistent with equilibrium thermodynamics in this case provided $u_\mathrm{xs}(\vc x)/T$ is small so that the exponential can be linearized, in which case the last term in eq.~(\ref{eq:gist}) may be written,
\begin{equation}
    T \ln \left \langle \exp \left (-\frac {u_\mathrm{xs}(\vc x)} {T} \right ) \right \rangle \approx -\langle u_\mathrm{xs}(\vc x) \rangle.
\end{equation}
For higher-order GIST, this consistent result extends also beyond the linear order of the Boltzmann factor.

\subsubsection{Short comment on 3D-2PT theory}
As for Ref.~\onlinecite{persson2017signatures}, the local entropy is computed using the 2PT theory of Lin and coworkers \cite{lin2003two}, and its extension to a discrete spatial grid is referred to as ``3D-2PT''. In this theory, the entropy is related to the dynamical properties of the fluid. Their sampling requires that the trajectories of the molecules be followed for a certain time, which however leads to a problem of precisely ascertaining their position in space and of ascribing the computed entropy solely to that position, however determined. Moreover, because the entropy depends on dynamic properties, the derivative with respect $N_i$ is especially intractable analytically. The question as to whether the local entropy function of Ref.~\onlinecite{persson2017signatures} satisfies eq.~(\ref{eq:def}) thus remains open, albeit finding that it does seems \textit{a priori} highly unlikely, given the very stringent requirements.

\section{Consistent spatial decomposition of the entropy}
\label{sec:consistent}
As shown in the previous section (with the exception of the spatially resolved 2PT theory, although the same conclusion is strongly suspected), the spatial decompositions of the energy and of the entropy found in the literature are in general rigorously inconsistent with each other in the sense that one and the same general definition for spatial resolution is not applicable to both thermodynamic functions, although the actual discrepancies might be numerically small and, indeed, for the first-order GIST, the definition is satisfied if one avoids phase transitions in the solvent phase diagram. Let us now turn to a simple alternative for the spatially resolved entropy function which is also consistent with eq.~(\ref{eq:def}) as well as with the ideal gas in an external field. Like this we prove constructively that even when the entropy decomposition is consistent with the energy one, there is still arbitrariness remaining in how to define it.

From the definitions of the Gibbs energy, $G$, and the chemical potential, we may write the local solvent entropy density as,
\begin{equation}
\label{eq:consistent-entropy}
\overline S(\vc x) = \frac {\overline H(\vc x)} T - \frac {\overline \mu(\vc x)} T 
\end{equation}
where $\overline \mu(\vc x) = \rho(\vc x) \left (\partial G / \partial N(\vc x) \right )_{T,P}$ is the chemical potential density of the solvent, and $\overline H(\vc x)$ is the solvent enthalpy density. This definition ensures that $\overline S(\vc x)$ satisfy eq.~(\ref{eq:def}), it being equivalent to 
\begin{equation}
\label{eq:entropy-consistent}
\overline S(\vc x) = \rho(\vc x) \left (\frac {\partial (H / T - G / T)} {\partial N(\vc x)} \right )_{T,P}.
\end{equation}
Neglecting the difference between the energy $U$ and enthalpy $H$ (which in condensed phases typically amounts to a negligibly small difference of a few joules per mole), we may write
\begin{equation}
\label{eq:entropy-approx}
\overline S(\vc x) \approx \frac {\overline U(\vc x)} T + \frac {3} {2} \rho(\vc x) - \frac {\overline \mu} T
\end{equation}
where $\overline U(\vc x)$ is computed as in Section~\ref{sec:energy}. The approximation, although very good for liquids and solids under ambient conditions, of substituting the energy for the enthalpy, is mandated by the problem of spatially resolving the pressure-volume term, a problem that remains without suggestions so far in the literature.

Provided the solute lacks internal degrees of freedom, eq.~(\ref{eq:consistent-entropy}) is simple to apply directly. In this way, we may write expressions for the entropy and energy of solvation consistent with each other, through eq.~(\ref{eq:solvation}):
\begin{equation}
\label{eq:entropy-solv}
\Delta_\mathrm{solv} S = \int \mathrm d \vc x [\overline S(\vc x) - \overline S_\infty]
\end{equation}
\begin{equation}
\label{eq:energy-solv}
\Delta_\mathrm{solv} U = \int \mathrm d \vc x [\overline U(\vc x) - \overline U_\infty]
\end{equation}
where $\overline U(\vc x)$ is computed using eq.~(\ref{eq:consistent-energy}), and $\overline S(\vc x)$ is computed using eq.~(\ref{eq:consistent-entropy}). The bulk values (denoted by subscript $\infty$) are computed for spatial points far away from the solute within the liquid phase. Now, when  inserting eq.~(\ref{eq:entropy-approx}) into eq.~(\ref{eq:entropy-solv}), we obtain
\begin{multline}
\label{eq:gibbs-duhem}
    \Delta_\mathrm{solv} S = \\ \int \mathrm d \vc x \left [ \frac {\overline U(\vc x)} {T} - \frac {\overline \mu(\vc x)} {T} - \frac {\overline U_\infty} {T} + \frac {\overline \mu_\infty} {T}  + \frac 3 2 \{ \rho(\vc x) - \rho_\infty \} \right ]  = \\ \frac {\Delta_\mathrm{solv} U} {T} - \frac {\Delta_\mathrm{solv} \mu} {T},
\end{multline}
where the identifications made in the last equality should be obvious after taking into account eq.~(\ref{eq:energy-solv}). Seeing as the solute that we have considered lacks all internal degrees of freedom, $\Delta_\mathrm{solv} \mu$ may be interpreted as the solvent contribution to the free energy of solvation, as is evident by simple rearrangement of eq.~(\ref{eq:gibbs-duhem}). Now $\overline \mu(\vc x) = \rho(\vc x) \mu$ and $\overline \mu_\infty = \rho_\infty \mu$ where $\mu$ is the constant chemical potential. The solvent contribution to the free energy of solvation may hence be written,
\begin{equation}
\label{eq:final}
\Delta_\mathrm{solv} \mu = \left (\mu - \frac 3 2 T \right ) \int \mathrm d \vc x [\rho(\vc x) - \rho_\infty]
\end{equation}
in compliance with eq.~(\ref{eq:solvation}).

That eq.~(\ref{eq:final}) expresses the free energy of solvation as a functional of the liquid density, is the result of our assumptions of a monoatomic solute and solvent and is the explicit end result of the goal introduced in the opening paragraph of Section~\ref{sec:solv}. However, eq.~(\ref{eq:final}) is not exact, even for the structureless monoatomic solute within the meanfield approximation. The most obvious deficiency is our neglect of the pressure-volume term in eq.~(\ref{eq:entropy-approx}) which means that eq.~(\ref{eq:final}) is at best valid only at low pressures. Moreover, our neglect  of the Gibbs-Duhem equation is a second source of error. Whereas the chemical potential of the solvent is spatially constant at equilibrium, it is not conserved in the solvation process, invalidating  eq.~(\ref{eq:final}) except at high dilution where this change is negligible. Finally, eq.~(\ref{eq:final}) only gives the solvent contributions to the free energy of solvation and the numerical value of $\mu$ (for a given solvent) is fixed by choice of reference point for the solvent energy $U$ only without any regard for solute or solute-solvent energy terms.

With this in mind, we note that eq.~(\ref{eq:final}) could, for instance, be used to interpret relative contributions to the free energy of solvation of a biomolecule from different parts of the molecule, from the fluid density in the ``solvation layer''. Provided an unambiguous definition of the ``solvation layer'' can be established, different analogs could in this way be compared. But great care must be taken in any such analysis to make sure the analysis is physically meaningful. Exactly how such an analysis could proceed is beyond the scope of this note.

\section{Conclusion: What is gained by spatial resolution?}
The experimental thermodynamic observables correspond to the definite integral in eq.~(\ref{eq:decomposition}), that is the total $A$, and are indifferent to the precise choice of the decomposition function $\overline A(\vc x)$ of which there is an infinite number. Ideally, spatial decomposition offers possible computational advantages in that a single molecular simulation -- to be more precise, a single and physical $6N$-dimensional Hamiltonian corresponding only to the intermolecular potential and kinetic energies -- may capture both bulk and solvation properties from different spatial regions of one and the same physical simulation, thus negating the need to run separate simulations (or use interpolating Hamiltonians with non-physical degrees of freedom) to compute the difference in $A$.

However, if the spatial variations of $\overline A(\vc x)$ themselves are to be interpreted in a physically meaningful way, it ought to be clear to what they correspond. But this is not the only concern: whereas GIST leads to a computed chemical potential that is constant at equilibrium, that obtained by GCT is not, in contradiction with the postulates of thermodynamics, unless a different way of spatially resolving the energy is chosen. However, taking the spatially resolved energy as anything other than the local binding energy would seem to lead to highly contrived analyses if one extends the argument to electronic degrees of freedom.

Nevertheless, although I have argued for maintaining consistency in the way entropy and energy are spatially resolved, and showed the nice -- and \emph{consistent} -- properties exhibited by the definition of eq.~(\ref{eq:def}), it is clear from the derivation in Section~\ref{sec:consistent} that the self-consistency requirement for the chemical potential amounts at most to a necessary, and not a sufficient, physical condition to define the spatial resolution. Therefore, like Mark and van Gunsteren \cite{mark94} before me, I still strongly caution against the overinterpretation of results from any molecular simulations that report spatially resolved thermodynamic functions, regardless of the way that the spatial resolution is obtained. This will remain the state of affairs until we have enough physical conditions to impose to make such an analysis unique and hence edifying.


\bibliography{paper}

\end{document}